# Interplay between superconductivity and ferromagnetism in epitaxial Nb(110)/Au(111)/Co(0001) trilayers

Hiroki Yamazaki,[1,*] Nic Shannon,[2] and Hidenori Takagi[1,3]

[1]*RIKEN Advanced Science Institute, 2-1 Hirosawa, Wako-shi, Saitama 351-0198, Japan*
[2]*University of Bristol, Tyndall Avenue, Bristol BS8 1TL, UK*
[3]*Department of Advanced Materials Science, Graduate School of Frontier Sciences, University of Tokyo, 5-1-5, Kashiwanoha, Kashiwa, Chiba 277-8851, Japan*

Epitaxially grown multilayer systems offer the possibility to study the influence of ferromagnetism on superconductivity in a new and controlled way. In this paper, we explore how the superconducting properties of high quality, epitaxially-grown superconductor/normal metal/ferromagnet trilayers evolve as a function of the exchange splitting in the ferromagnet, and the thickness of the normal metal layer. We report results for Nb(110)/Au(111)/Co(0001), and make a detailed comparison with earlier results for Nb(110)/Au(111)/Fe(110). We use quantitative FFT analysis to confirm the existence of a long-period (2.1 nm) oscillation in the superconducting transition temperature $T_c$ as a function of the Au-layer thickness $t_{Au}$, for $t_{Au}>2$ nm, and highlight an additional short-period (0.76 nm) oscillation for $t_{Au}<3$ nm in Nb/Au/Co. This short-period oscillation can be explained in terms of a damped RKKY-like oscillation of the spin-polarization in Au. The robustness of the long-period oscillation against the substitution of Co for Fe suggests that it is intrinsic to the Au(111) layer on Nb, and may represent a new form of quantum interference in very clean trilayer systems.

# I. INTRODUCTION

The properties of artificially fabricated superconductor/ferromagnet (SC/FM) junctions have been studied intensively in the last decade. The interest of these systems lies in the possibility of mixing superconducting and ferromagnetic correlations in a controlled way. In particular, the proximity effect allows Cooper pairs to penetrate from the SC into the FM, where they experience a large exchange field. Predictions that this should lead to Fulde-Ferrel-Larkin-Ovchinnikov (FFLO) oscillations in the superconducting order parameter,[1,2] were confirmed spectacularly by experiments on SC/FM/SC trilayers.[3]

In recent years, a number of authors have also investigated the experimental[4-9] and theoretical[10,11] properties of SC/NM/FM trilayer systems, in which a layer of normal metal (NM) is introduced between a SC and a FM. The interaction between SC and FM is now mediated by the conduction electrons of the intermediate NM layer. These electrons are subject to both proximity-induced superconductivity *and* exchange-induced spin polarization, and carry additional information about the band structure and thickness of the NM layer. High quality, epitaxially grown SC/NM/FM multilayers therefore offer a unique opportunity to investigate the properties of superconducting electrons under the influence of an exchange field in a controlled manner.

Two important length scales control the way in which SC and FM combine in the NM. The first is "the normal-metal coherence length" $\xi_N$ over which proximity-induced superconductivity at the NM/SC interface decays in the NM. The second is "the spin penetration depth" $\xi_{sp}$ over which spin polarization induced at the FM/NM interface decays in the NM. These length scales are illustrated schematically in Fig. 1. Pioneering studies of Fe/Pt/Nb multilayers[4] demonstrated that the range of the interaction between Nb and Fe is limited not by $\xi_{sp}^{Pt}$ (~a few atomic planes), but rather by $\xi_N^{Pt}$~32 nm($\gg\xi_{sp}^{Pt}$). Accordingly, when the thickness of the NM layer ($t_{NM}$) is less than $\xi_{sp}^{NM}$ ($t_{NM}<\xi_{sp}^{NM}$; see Fig. 1(a)), the SC layer comes in direct contact with the FM-induced spin polarization of the conduction electrons at the SC/NM interface, and the superconducting transition temperature ($T_c$) is suppressed. This spin polarization is accompanied by the effective magnetic exchange field ($h_{ex}$) and is expected to exhibit a Ruderman-Kittel-Kasuya-Yoshida (RKKY) oscillation analogous to that observed in some FM/NM/FM trilayers.[12]

On the other hand, when $t_{NM}$ takes a value between $\xi_{sp}^{NM}$ and $\xi_N^{NM}$ ($\xi_{sp}^{NM}<t_{NM}<\xi_N^{NM}$; see Fig. 1(b)), we enter a regime where *indirect* coupling between SC and FM can occur within the NM layer. This situation raises the possibility of new quantum coherence effects arising from the competition between superconductivity and ferromagnetism within the NM layer, especially in clean, high quality, epitaxially grown SC/NM/FM trilayers. How this competition is resolved in a SC/NM/FM trilayer is an open question, and one which poses a significant challenge to theory and experiment alike.

In our published work on Nb/Au/Fe trilayers,[7] we confirmed the suppression of $T_c$ observed in the first experiments on Fe/Pt/Nb trilayers,[4] and identified a structural transition and a long-period oscillation in $T_c$ as a function of the Au layer thickness ($t_{Au}$). At the same time, similar findings were presented by Kim *et al.* for Nb/Au/CoFe trilayers.[5] However the period of the long-wavelength $T_c$ oscillation was different in each case, and neither study reported the short-period oscillation in $T_c$ which might have been anticipated from RKKY interactions known to be present in FM/NM multilayers.[12] These papers posed a number of important questions, which current theory of multilayers simply does not address. What is the mechanism of the long-period $T_c$ oscillation? What is the role of the structural transition? Why were RKKY oscillations not observed? How would the period of these oscillations be affected by the substitution of a different FM layer with different exchange field?

In the present paper, we address these questions by fabricating SC/NM/FM trilayers of high quality using a different ferromagnetic metal. In these new Nb/Au/Co trilayers, the structural transition related to the Au/Fe lattice mismatch is eliminated, and we are able to observe, for the first time, the short-period oscillations in $T_c$ associated with RKKY couplings. We use a quantitative fast Fourier transform (FFT) analysis to confirm the existence of the long-period (2.1 nm) oscillation previously reported for Nb/Au/Fe trilayers, and further demonstrate that this oscillation is robust against the substitution of Co for Fe. The period of this oscillation is therefore independent of the exchange splitting in the FM layer. The question of how to correctly identify the superconducting transition temperature from susceptibility measurements has also been resolved by studying the resistive transition. These results present a serious challenge to the developing theory of SC/NM/FM trilayers, and we hope that they will also stimulate further careful experimental study.

This paper is organized as follows. In Sec. II, details of the sample preparation are presented. The structural properties of the samples are discussed based on the results of X-ray diffraction studies (Sec. III). Section IV focuses on the magnetic and superconducting properties of the samples, in particular the dependence of $T_c$ on the thickness of the Au layer is eventually presented. Section V is devoted to the FFT analysis of the $T_c(t_{Au})$ data. In Sec. VI we discuss the possible mechanisms which might drive the oscillation in $T_c(t_{Au})$. We conclude with brief summary of results and open questions in Sec. VII.

**II. SAMPLE PREPARATION**

A series of Nb(110)/Au(111)/Co(0001) trilayers were prepared on a substrate of single crystal $Al_2O_3(11\bar{2}0)$ using a molecular-beam-epitaxy (MBE) machine (Eiko Co., Japan). Details of the sample preparation were essentially the same as those for Nb(110)/Au(111)/Fe(110) [Ref. 7], and considerable care was taken to reproduce the same preparation conditions for different samples. The thickness of the Nb layer (28.8 nm) was chosen to be of the same order as the superconducting coherence length $\xi(0) \sim 41$ nm (Ref. 13) of bulk Nb, to achieve a clean-limit SC at absolute zero. A Au layer of thickness $t_{Au}$=0.00-13.05 nm (0-55 ML), and a Co layer of thickness 12.6 nm (~62 ML) were deposited at a rate of 0.010±0.005 nm/sec. The trilayer was finally capped with a Au layer of 4.4 nm (~19 ML) in order to avoid oxidization of the Co layer.

During the sample growth, reflection high-energy electron diffraction (RHEED) patterns were measured for the surface of *each* layer. Fine streak patterns (Fig. 2) were obtained for all of the layers, demonstrating epitaxial growth within each layer. For the Au layer [Fig. 2(b)], the pattern and the interval between the streaks were independent of $t_{Au}$ from 0.44 nm (~2 ML) to 13.05 nm (~55 ML), confirming epitaxial layer-by-layer growth of Au on top of the Nb.  We also find that the Co layer grows epitaxially on the Au layer even in the presence of a large (~14%) misfit between hcp Co(0001) and fcc Au(111).[14]  In the case of the Nb(110)/Au(111)/Fe(110) trilayers, the RHEED spectra for the Fe layer showed two patterns of single domain ($t_{Au}$<1.7 nm) and twinned domain ($t_{Au}$≥1.7 nm) structures.[7]  In contrast, the RHEED pattern of the Co layer in the new Nb/Au/Co trilayers exhibits the pattern shown in Fig. 2(c) for *all* values of $t_{Au}$.  This implies that the Co layer maintains an identical crystal structure regardless of the thickness of the Au layer.

**III. STRUCTURAL CHARACTERIZATION**

*Ex situ* structural characterization was performed mainly by x-ray diffraction measurements using the x-ray diffractometer system of Philips MRD with Cu $K\alpha_1$ radiation. A typical reflection pattern of the middle-angle 2θ-θ scan is shown in Fig. 3 for $t_{Au}$=2.2 nm. We see the presence of well-resolved Laue oscillations from Nb as well as Co.  The degree of coherence of crystal growth in a given layer can be estimated from the width of the corresponding peak in the 2θ-θ scan.  All the samples show bcc Nb(110) and hcp Co(0001) crystal growths with a single domain covering the *full thickness* of the Nb and Co layers. The rocking curves of the Nb(110) and Co(0001) Bragg peaks have typically a full width at half maximum of 0.05° and 0.7°, respectively.  The Au(111) peak is not well resolved at intermediate angles because of the small separation of the Nb(110) and Au(111) peaks, and of the broadening of the Au(111) peak due to a small thickness of the Au layer.  However the Au(222) peak at a high angle of 2θ~82.4° can be resolved from the Nb(220) peak for larger $t_{Au}$'s.

The trilayer's crystal orientation in the sample plane was determined from x-ray diffraction φ-scans, where φ is the sample rotation angle around the axis perpendicular to the sample plane.  During the scans, the angles of 2θ and ω were fixed to the Nb(310), Au(311), and Co(10$\bar{1}$3) Bragg conditions, so that scattering vectors are not perpendicular to the sample plane (see the inset of Fig. 4).  Typical diffraction patterns are shown in Fig. 4 for $t_{Au}$=3.9 nm.  The Nb(310) peaks exhibit twofold symmetry in φ showing good single-crystal growth of the Nb(110) layer on the substrate.  The Au(311) peaks show alternate change of intensity; i. e., the peaks at −150°, −30°, and 90° are stronger than the others. This is likely due to twinned fcc structure (two fcc single-crystal domains twinned by 60°) of the Au(111) layer.  We could not distinguish the peaks of Au in the trilayer from those of the cap Au layer.  The presence of the (10$\bar{1}$3) peaks of Co (2θ=84.33°, ω=9.89°) proves the Co layer of hcp and not of fcc structure because there is no corresponding fcc peak at this position.  The sixfold symmetry of these peaks naturally comes from the hcp Co(0001) layer.  We confirm that the Co peak at ~44.4° in the 2θ-θ scan (Fig. 3) is not the peak of fcc Co(111) but of hcp Co(0001), although they have close *d* values.  Since the direction of φ=0° is parallel to the <0001> axis of the Al$_2$O$_3$(11$\bar{2}$0) substrate, we conclude Nb<$\bar{1}$11>// Au<$\bar{1}$10>//Co<11$\bar{2}$0>//Al$_2$O$_3$<0001>.

In order to check the roughness of interfaces, the diffraction intensity measured in the small-angle 2θ-θ scans was fitted using x-ray diffraction profile program SUPREX (developed by Fullerton et al.[15]). Typical diffraction data for $t_{Au}$=2.8 nm (circles) and the result of fitting (solid curve) are shown in Fig. 5. The experimental data are well reproduced by a surface and interface roughness parameter $\sigma_n$ (n=0-4) (inset of Fig. 5) and the known layer thicknesses (inset of Fig. 3). Here the distribution of dϱ(z)/dz in the vicinity of the interface is regarded as a Gaussian profile with a standard deviation of $\sigma_n$, where ϱ(z) is the electron density as a function of distance z measured perpendicular to the interface. We see that the trilayer has a very sharp interface at the bottom of the Au layer ($\sigma_3$=0.00 nm). However, the value of $\sigma_2$=0.44 nm (~2 ML) associated with the Au/Co interface is larger than the corresponding value of $\sigma_2$=0.19 nm (<1 ML) for the Au/Fe interface in our earlier Nb/Au/Fe trilayers.[7] We note that, while Au and Co are immiscible in the bulk, a partial exchange of Au and Co atoms may occur at an interface.[16, 17] The substantial difference in values of $\sigma_2$ found for Au/Co and Au/Fe is probably due to the different initial growth modes of Co and Fe on the reconstructed Au surface: Co islands show a bilayer-growth,[18] in contrast to the monolayer-growth of Fe.[19] The formation of an interface-confined mixture at the Au/Co interface will effectively decrease $t_{Au}$. The consequences of a decrease of $t_{Au}$ by ~2 ML are discussed further below in the context of the $t_{Au}$ dependence of the superconducting transition temperature $T_c$.

**IV. MAGNETIC AND SUPERCONDUCTING PROPERTIES**

In order to investigate the magnetic and superconducting properties of these trilayers, magnetization and magnetic susceptibility measurements were performed using a superconducting quantum interference device magnetometer (Quantum Design MPMS2). Resistivity measurements were carried out in a standard four-terminal configuration using low-frequency ac technique with a current density less than 1 A/cm$^2$. The current and voltage leads were attached to the sample surface by silver epoxy.

The magnetic anisotropy of the Co layer was determined by measuring magnetic hysteresis loops. Typical results at 8.0 K (>$T_c^{on}$=7.82 K) are shown in Fig. 6 for $t_{Au}$=8.7 nm. A magnetic field H was applied parallel (solid curve) and perpendicular (dashed curve) to the sample plane. The loop exhibits a coercive field of ~30 Oe for H // sample plane. The figure shows an apparent in-plane anisotropy, and this can be confirmed for all the Nb/Au/Co trilayers. The greater part of the Co magnetization lies in the sample plane; a small perpendicular component with hysteresis between −1 and 1 kOe is also seen, but this amounts to less than 10% of the saturation magnetization. This is a small perpendicular component, and within experimental accuracy this shows no dependence on $t_{Au}$. The strong in-plane anisotropy can easily be understood in terms of the shape anisotropy of the ferromagnetic film.

Conducting and superconducting properties of a single Nb layer and Nb/Au bilayers have been already investigated in the context of Nb/Au/Fe trilayers.[7] Here we summarize the results below. A single Nb layer of 28.8 nm (without any capping layer on it) exhibits $T_c$=9.04±0.01 K (at 10% of the normal-state resistivity) with a transition width of 0.04±0.01

K (10-90% criterion). The effective mean free path of electrons ($l_{\text{eff}}^{\text{Nb}} \sim 25$ nm) and the Ginzburg-Landau coherence length at absolute zero [$\xi_{\text{GL}}(0) \sim 32$ nm] are comparable with the Nb-layer thickness of 28.8 nm. The residual resistivity of the Nb/Au bilayers was measured at 9.1 K ($>T_c$). The typical resistivities of the Nb and Au layers were estimated to be $\rho_n^{\text{Nb}} = 1.5 \times 10^{-8}$ $\Omega$m and $\rho_n^{\text{Au}} = 3.3 \times 10^{-9}$ $\Omega$m. The measured value of $\rho_n^{\text{Au}}$ corresponds to a mean free path of $l^{\text{Au}} \sim 250$ nm [Ref. 7], which is sufficiently larger than the thicknesses of the Au layer. Charge transport through the Au spacer layer is therefore essentially ballistic. It has also been confirmed that the superconducting order parameter extends throughout the Au layer, i. e., $t_{\text{Au}} \leq \xi_N^{\text{Au}}$ holds at least up to $t_{\text{Au}} \sim 10$ nm.[7]

Typical temperature dependences of the normalized magnetic susceptibility $\chi_n = \chi / |\chi_{(2K)}|$ and of the normalized resistance $R_n = R/R_{(9K)}$ are shown in the inset of Fig. 7 for the trilayer of $t_{\text{Au}} = 2.6$ nm, where $\chi_{(2K)}$ is the susceptibility at 2 K and $R_{(9K)}$ is the resistance at 9 K. The susceptibility measurements were carried out for the warming (after zero-field cooling in $|H| < 0.002$ Oe) and cooling procedures in a magnetic field of 2 Oe applied perpendicular to the sample plane. A clear indication of superconductivity can be seen in the diamagnetic response below $T_c^{\text{on}}$, which agrees with the value of $T_c$ estimated by zero-field resistivity measurements within experimental accuracy. Resistivity measurements show that the trilayers exhibit the same transition width of $0.04 \pm 0.01$ K (10-90% criterion) as observed for a single Nb layer. The specious transition width of $\sim 1$ K in susceptibility is therefore mainly due to the flux penetration for a small thickness of the Nb layer (28.8 nm) less than the penetration depth $\lambda(0)$ of 39 nm [Ref. 20]. In this paper, $T_c^{50\%}$ is defined as the temperature at which $\chi$ has a 50% value of $\chi_{(2K)}$. The dependence of $T_c$'s ($T_c^{\text{on}}$ and $T_c^{50\%}$) on $t_{\text{Au}}$ is shown in Fig. 7 together with the $T_c^{\text{on}}$ of the Nb/Au bilayers[7] for comparison. The solid curve shows fits to the bilayer data using the McMillan expression described in Ref. 7. As $t_{\text{Au}}$ decreases below $\sim 3$ nm, the difference ($T_c^{\text{on}} - T_c^{50\%}$) increases slowly, but does not show a sudden jump like that seen in the Nb/Au/Fe trilayers at $t_{\text{Au}} = 1.7$ nm.[7] This result, together with the results of RHEED, confirms the absence of structural change in the Nb/Au/Co trilayers as a function of $t_{\text{Au}}$.

The $T_c^{\text{on}}$ of the Nb/Au bilayers decreases monotonically with $t_{\text{Au}}$, exhibiting the usual superconducting proximity effect, while that for the Nb/Au/Co trilayers exhibits completely the opposite behavior: $T_c^{\text{on}}$ is most strongly suppressed for $t_{\text{Au}} = 0$ nm, and (on average) increases with increasing $t_{\text{Au}}$, suggesting that the Au layer partially screens the superconducting Nb from the ferromagnetic Co. The markedly different values of $T_c^{\text{on}}$ for Nb/Au and Nb/Au/Co are strong evidence that superconductivity couples with ferromagnetism for $t_{\text{Au}}$'s at least up to 10 nm. At $t_{\text{Au}} = 0$ nm, where there is direct contact between Nb and Co, $T_c^{\text{on}}$ is reduced by 2.74 K relative to that of a single layer of Nb (i.e. the Nb/Au bilayer with $t_{\text{Au}} = 0$ nm) — a change of $\sim 30\%$. This suppression of $T_c^{\text{on}}$ should be compared with a change of $\Delta T_c^{\text{on}} = 3.70$ K for the Nb/Fe bilayer.[7] Comparing these values with experimental estimates of the exchange splittings ($\delta E_{\text{ex}}$'s) of 1.0 eV in Co and 1.5 eV in Fe,[21] we find that the reduction of $T_c^{\text{on}}$ due to direct proximity of a FM is roughly proportional to $\delta E_{ex}$.

For $t_{Au}<3$ nm, we clearly see an oscillating change in $T_c^{on}$ and $T_c^{50\%}$ with a (short) period of 0.76 nm (~3.2 ML of Au), superimposed on the steep overall rise of $T_c(t_{Au})$. For clarity, vertical broken lines are drawn at intervals of 0.76 nm in Fig. 7. The period and amplitude (about 7% of $T_c$) of the oscillation remain clearly defined up to $t_{Au}\sim 3$ nm, for both $T_c^{on}$ and $T_c^{50\%}$. In addition to these short-period oscillations, we can distinguish some quasiperiodic local maxima and minima of $T_c$'s for $t_{Au}>2$ nm. While it would be hard to unambiguously fit a single period to this structure, it is interesting to plot data from Nb/Au/Co trilayers together with equivalent measurements of the Nb/Au/Fe trilayers.[7] This is done in Fig. 8, which suggests that long-period (2.1 nm~9 ML) oscillations in $T_c(t_{Au})$ *do* occur for both compositions, at least for 2 nm$<t_{Au}<$10 nm. A quantitative FFT analysis, confirming the period of these oscillations, is given below. We note that in Fig. 8, $t_{Au}^{eff}$ is the effective thickness of the Au layer, calculated as $t_{Au}^{eff}=t_{Au}$ for Nb/Au/Fe and $t_{Au}^{eff}=t_{Au}-0.44$ nm for Nb/Au/Co. This correction compensates for the mixing of Au and Co atoms at the Au/Co interface (see discussion in Sec. III), and allows us to directly compare results for Nb/Au/Co and Nb/Au/Fe.

Remarkably, within experimental accuracy, the Nb/Au/Co and Nb/Au/Fe trilayers have the same values of $T_c^{on}$ and $T_c^{50\%}$ on the dashed lines at $t_{Au}^{eff}=n\times 2.1$ nm ($n$: integers) — except for $n=0$ and $n=1$. Between these dashed lines, although less evident for $4.2<t_{Au}^{eff}<6.3$ nm, the superconducting transition temperature $T_c$ for Nb/Au/Co (Nb/Au/Fe) is a concave (convex) function of $t_{Au}^{eff}$. The $T_c$'s for Nb/Au/Co are therefore *always* lower than those of Nb/Au/Fe.

## V. APPLICATION OF FFT

In order to confirm that the $T_c^{on}(t_{Au}^{eff})$ and $T_c^{50\%}(t_{Au}^{eff})$ data have a recognizable oscillation period of 2.1 nm, a more quantitative analysis of the data is required. The power spectra of the data were therefore calculated for Nb/Au/Co and Nb/Au/Fe (see Fig. 9). The Fast Fourier Transform (FFT) was applied to the data of $t_{Au}^{eff}>4$ nm ($N=20$) for Nb/Au/Co and to those of $t_{Au}^{eff}>2$ nm ($N=20$) for Nb/Au/Fe, where $N$ is the number of data points. The interval in $t_{Au}^{eff}$ is 0.435 nm($=\Delta t_{Au}^{eff}$), and the horizontal scaling in the figure represents $n/(N\cdot \Delta t_{Au}^{eff})$. Since the data points are not perfectly evenly spaced for all the Nb/Au/Fe data, we have compensated for the data points at $t_{Au}^{eff}=4.350$ and 4.785 nm using a linear interpolation, and have adopted the data points of a multiple of 0.435 nm in $t_{Au}^{eff}$.

The intensity of $P$ at $n=0, 1$ is not indicated in Fig. 9 for clarity. The key point is the peak at $n=4$, which corresponds to an oscillation with a period of $(N\cdot \Delta t_{Au}^{eff})/4=2.175$ nm. This peak can be clearly recognized not only for Nb/Au/Fe but also for Nb/Au/Co, suggesting the presence of an (incommensurate) oscillation period of ~2.2 nm in both cases. We have checked the stability of the peak at $n=4$ under modification of the original $T_c(t_{Au}^{eff})$ data by applying a window function. A typical window function in the FFT operation, such as Hamming or Hanning one, was applied to the data set, and the results of FFT still showed a peak at $n=4$. It was also confirmed that the peak that corresponds to an oscillation period of ~2.2 nm is stable under reduction in $N$ from 20 to 14; this was done by omitting gradually the $T_c(t_{Au}^{eff})$ data points from the left-side end (smallest $t_{Au}^{eff}$) or from the right-side end (largest $t_{Au}^{eff}$). These results indicate that the peak at $n=4$ is a true peak, and other shorter lines are the power spectrum of the noise. In fact, local peaks such as one at $n=7$ for $T_c^{on}$ of Nb/Au/Co are unstable under the reduction in $N$ mentioned above. We conclude that the $T_c^{on}(t_{Au}^{eff})$ and $T_c^{50\%}(t_{Au}^{eff})$ data have a recognizable oscillation period of ~2.2 nm very close to the long-period of 2.1 nm claimed in the former section for both Nb/Au/Co and Nb/Au/Fe. To the best of our knowledge, this is the first time that such a data set for a multilayer system has been analyzed in this way. For shorter data sets, a better approach might be to perform an unbiased maximum entropy method to fit the data. However for the data sets considered here, FFT analysis yields satisfactory results.

One remaining ambiguity is the possibility of "aliasing", the possible confusion of short- and long-period signals, as illustrated in Fig. 10. When sampled only at discrete positions of $L=n\times0.435$ nm ($n$: integers), both the periodic functions with periods of 0.243(±0.001) and 0.549(±0.007) nm cannot be distinguished from a periodic function with a period of 2.1(±0.1) nm, where the error in the short period was estimated on the basis of the error (±0.1) for the long period. Thus from FFT alone we cannot completely rule out the possibility that the long-period oscillations observed could be due to an underlying short-period oscillation. However, the data points at $t_{Au}^{eff} \neq n\times0.435$ nm, which were excluded from this FFT analysis (specifically points for $t_{Au}^{eff}$=2.436, 3.306, 6.786, and 10.222 nm for Nb/Au/Fe) seem to exhibit the same long-period oscillation as the data used for the FFT, and show no hint of a shorter period (cf. Fig. 8). Considering all of these facts together, we are therefore confident that the long-period (2.1 nm) oscillations are a robust, intrinsic effect in these trilayer systems.

**VI. DISCUSSION**

The main challenge posed by current experiments on SC/NM/FM trilayers is to understand the mechanism(s) which determine the superconducting transition temperature $T_c$, as a function of the thickness of the normal metal layer NM.[7,8] The comparative study of Nb/Au/Co and Nb/Au/Fe trilayers presented here makes it possible to put this question on a firmer footing. Three key features emerge: (1) "long-period" oscillations with a period of 2.1 nm (~9 ML) are seen for $t_{Au}$>2 nm, while for $t_{Au}$<3 nm "short-period" oscillations with a period of 0.76 nm (~3.2 ML) are seen in Nb/Au/Co trilayers but were *not* observed for Nb/Au/Fe trilayers. (2) The long-period oscillation is robust against the substitution of Co for Fe. (3) After correction for interface effects, we find that the long period oscillations always have $T_c^{Nb/Au/Co} \leq T_c^{Nb/Au/Fe}$ and are apparently in antiphase between the two sets of trilayers.

In our previous papers,[7,8] we explored possible routes to a long-period oscillation in $T_c$, namely (a) a periodic Friedel-type oscillation in the density of electrons in the NM layer, (b) FFLO oscillations in the superfluid density, and (c) an RKKY type of oscillation in the effective magnetic exchange field across the NM layer. None of these was found to offer a satisfactory explanation of our results for Nb/Au/Fe trilayers. However (c) offers a very plausible explanation of the short-period oscillation observed in Nb/Au/Co.

RKKY couplings in FM/NM multilayer systems have been extensively studied in the context of giant magneto-resistance. The effective exchange coupling between neighboring FM layers is found to oscillate (and in fact to change sign) as a function of the thickness of the normal metal "spacer" layer NM.[12, 22] This effect can be understood in terms of a (damped) oscillation of the effective exchange field ($h_{ex}$) felt by conduction electrons in the NM layer, induced by the exchange field of the FM. In our trilayers, where the thickness of the NM layer $t_{Au}$ is much less than the penetration depth of the spin polarization induced in the Au layer ($\xi_{sp}^{Au}$), superconductivity in the Nb layer is suppressed by direct contact with a large $h_{ex}$ at the Nb/Au interface [similar to the case in Fig. 1(a)]. As $t_{Au}$ ($<\xi_{sp}^{Au}$) increases, $h_{ex}$ at the Nb/Au interface exhibits an RKKY like oscillation with a period determined (up to a reciprocal lattice vector) by the nesting vectors of the Fermi surface of the normal metal in the direction *perpendicular* to the NM/SC interface. Since the exchange field is pair-breaking for either sign of exchange, this would lead to an oscillation in $T_c$ with *half* the period of the RKKY oscillation in the exchange field. Direct evidence for such an RKKY mechanism of $T_c$ oscillations may be possible through the use of μSR or β-NMR appropriate depth resolution to detect oscillations in the local magnetic exchange field of thin films of Au on Co — indeed the coexistence of ferromagnetism and superconductivity in FM/SC/FM trilayers has recently been studied using low energy μSR.[23]

From this picture we can obtain a good qualitative — and even semiquantitative — understanding of the short-period $T_c$ oscillation observed for small $t_{Au}$'s in Nb/Au/Co trilayers. The range of thickness (~3 nm) for which short-period oscillations are observed is of the same order of magnitude as the spin penetration depth $\xi_{sp}^{Au} \approx 1$ nm found from Mössbauer spectroscopy of Co/Au multilayers.[24] And the observed period of 0.76 nm is close to the theoretical prediction (1.14 nm)/2=0.57 nm found from the relevant nesting vector of bulk Au, allowing for umklapp scattering.[22] The small (~25%) discrepancy could plausibly originate in a modification of the Fermi surface by small changes in the lattice parameters of Au when it is grown epitaxially on Nb. The absence of an observable short-period oscillation in the Nb/Au/Fe trilayers need not contradict this model, since the amplitude of $h_{ex}$ is smaller in a Au layer when it is adjacent to Fe than to Co.[25] The picture for Nb/Au/Fe trilayer is further complicated by a structural change in the Fe layer as a function of $t_{Au}$,[7] which leads to a jump in $T_c$ for $t_{Au}$=1.7 nm, making the identification of any short-period $T_c$ oscillation very difficult.

The most striking finding of this study is the coincidence of $T_c$'s in Nb/Au/Co and Nb/Au/Fe especially for $t_{Au}^{eff}=n\times 2.1$ nm (where $n>1$ is an integer), despite the fact that both the exchange splitting $\delta E_{ex}$ of the FM layer, and the interface resistance $\gamma^{NM/FM}$ at the NM/FM interface are very different. It is also interesting to note that we can see the robustness of the long-period (2.1 nm) oscillation in $T_c$ against the substitution of Co for Fe in the trilayers. These results provide strong evidence that, for $t_{Au}>2$ nm, the overall behavior of $T_c(t_{Au})$ and the long-period oscillation (except for its phase) are intrinsic to the Au(111) layer on Nb. In the meantime, it is also certain that the weak interplay between the FM layer and the Nb layer remains even for $t_{Au}>2$ nm, since the $T_c$'s of Nb/Au/Co and Nb/Au/Fe (Figs. 7 and 8) are clearly lower than those of the Nb/Au bilayers (Fig. 7).

The challenge presented by these results is to establish a mechanism for $T_c$ oscillations in a SC/NM/FM trilayer which is independent of the exchange splitting $\delta E_{ex}$ and interface resistance $\gamma^{NM/FM}$ of the FM layer. This will necessarily involve constructing a theory of the interplay between SC and FM via NM for the $t_{NM}$'s beyond the RKKY-coupling range. An interesting development in this direction is the recent theory for conductance fluctuations as a function of NM thickness in FM/NM/SC trilayers.[26]

The long-period oscillations dominate for $t_{Au}$>2 nm, distances greater than the spin correlation length $\xi_{sp}^{Au}$≈1 nm. This means that the exchange splitting at the Nb/Au interface is negligible [a similar case to Fig. 1(b)], in marked contrast to the RKKY-like coupling at small $t_{Au}$ [a similar case to Fig. 1(a)]. A different, indirect coupling between SC and FM must occur within the Au layer. Since the charge transport through the Au spacer layer is essentially ballistic, it cannot be described only in terms of macroscopic quantities such as the exchange splitting of the FM $\delta E_{ex}$ and the boundary resistance of the junction $\gamma^{NM/FM}$ alone — quantum coherence effects must also be taken into account.

In particular, the Andreev reflection of quasiparticles at the SC/NM boundary, and their spin-polarized scattering at the NM/FM boundary, must be treated properly. Coherent multiple scattering of electrons between these two interfaces can lead to the formation of "particle-in-a-box" bound states with nontrivial boundary conditions. Quite subtle effects can follow; in particular, SC/FM interfaces with zero-energy Andreev bound states are unstable against a ground state with a spontaneous current.[27] The inclusion of a normal metal spacer provides a means of tuning bound states through the Fermi energy, and so of switching this spontaneous current on (and off). A recent theoretical treatment of a SC/NM/FM trilayer predicts that a persistent current flows in the SC/NM and NM/FM boundaries for spacer thicknesses which — in the simple model considered — are multiples of ~6 ML. This results in an oscillation in ground state energy with period ~6 ML [Ref. 10], of similar magnitude to the ~9 ML $T_c$ oscillations observed here. More work is needed to clarify whether this Andreev bound state mechanism is indeed effective in Nb/Au/Co and Nb/Au/Fe trilayers. However we note that both the magnetic field generated by the spontaneous current, and the modification of the density of states by the Andreev bound state are, in principle, observable.[27]

A possible explanation for the opposite phases of long-period $T_c$ oscillations in Nb/Au/Co and Nb/Au/Fe trilayers can be found in different character of Fe and Co bands at the Fermi energy: Fe has mostly majority-spin states, while Co has only minority-spin states. This will give rise to the bound states of minority spin (majority spin) in the Au layer when it is adjacent to Fe (Co),[28-30] and Cooper pairs will feel the opposite sense of exchange fields when traversing the Au/Fe (Au/Co) interface. Quite generally, spin dependent scattering at the NM/FM interface may open a route to more exotic forms of superconductivity, including triplet pairing.[31] The spin-flip scattering of Cooper pairs from magnetic atoms at the interface[32] might also explain why the FM with the large exchange splitting (Fe) is less effective in suppressing the $T_c$ of the superconducting state ($T_c^{Nb/Au/Co}$<$T_c^{Nb/Au/Fe}$ although $\delta E_{ex}^{Co} < \delta E_{ex}^{Fe}$). However their role in the periodic $T_c$ oscillations remains unclear.

Another unresolved issue is the role of spin-orbit coupling (SOC) in Au. Where SOC lifts the degeneracy between "up" and "down" spin electrons, Cooper pairs can form with a finite momentum, leading to FFLO-like oscillations of the superconducting order parameter in real space. SOC is particularly pronounced in Au(111) surface states, where a lack of inversion symmetry leads to a large splitting[33] [here a parallel can be made with non-centrosymmetric superconductors such as CePt$_3$Si (Ref. 34)]. However, even if these surface states survive in the SC/NM/FM system, they would *not* be expected to contribute substantially to electronic states *perpendicular* to the interface. This issue clearly merits further investigation.

## VII. CONCLUSIONS

In this paper we have explored the subtle interplay between ferromagnetism and superconductivity in artificial SC/NM/FM heterostructures, through a detailed experimental study of high quality Nb/Au/Co trilayers, and careful comparison with previous results for identically grown Nb/Au/Fe trilayers. A number of the questions raised in the preceding studies have been answered. A short-period (0.76 nm) oscillation of $T_c(t_{Au})$ attributed to the RKKY coupling was observed for the first time in Nb/Au/Co trilayers, which are free from a structural transition as a function of $t_{Au}$. We have further confirmed the existence of a long-period (2.1 nm) oscillation by quantitative FFT analysis of $T_c(t_{Au})$, and demonstrated that this oscillation is robust against the substitution of Co for Fe, and therefore independent of the exchange splitting of the FM layer. This long-period oscillation is interpreted as a new form of quantum interference effect, intrinsic to the Au normal metal layer. However the mechanism underlying this oscillation remains unclear, and these results motivate a thorough reexamination of the theory of SC/NM/FM trilayers. Further theoretical work, treating a full range of quantum coherence effects, is needed to clarify what novel bound states can exist as a function of the thickness of the normal metal layer $t_{Au}$, and their influence on the superconducting $T_c$. The phase inversion of the long-period $T_c$ oscillation when Co is substituted for Fe, and the apparent coincidence in values of $T_c$ for Nb/Au/Co and Nb/Au/Fe trilayers at $t_{Au}^{eff} = n \times 2.1$ nm (*n*: integers) provide further constraints on future theory.

**ACKNOWLEDGEMENTS**

The authors thank Prof. J. F. Annett and Prof. B. L. Györffy for many helpful comments on this study.

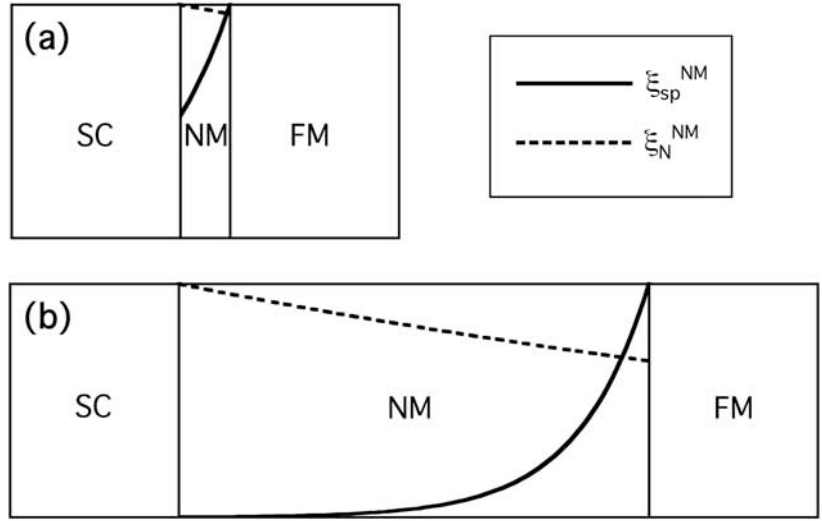

**FIG. 1.** Two length scales $\xi_{sp}^{NM}$ and $\xi_{N}^{NM}$ in the normal-metal layer of SC/NM/FM system. The proximity-induced pairing amplitude (dashed curve) decays with a length scale $\xi_{N}^{NM}$, while the spin polarization induced at the NM/FM interface (solid curve) decays with a length scale $\xi_{sp}^{NM}$. Assuming exponential decay, two cases are shown: (a) $t_{NM} < \xi_{sp}^{NM} < \xi_{N}^{NM}$ and (b) $\xi_{sp}^{NM} < t_{NM} < \xi_{N}^{NM}$, where $t_{NM}$ is the thickness of the NM layer. In our trilayer system, cases (a) and (b) are typical of the regimes for $t_{Au} < 3$ nm and for $t_{Au} > 2$ nm, respectively.

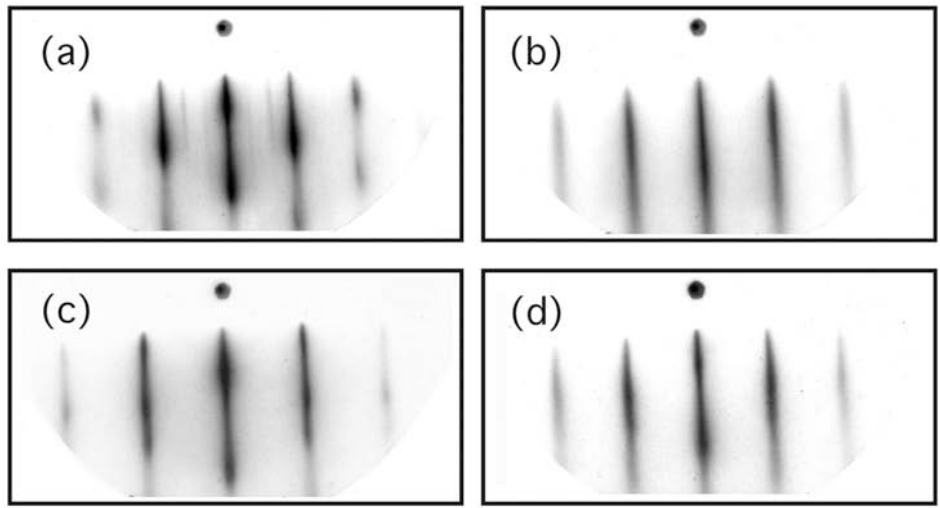

**FIG. 2.** Reversal images of typical RHEED patterns obtained in the growth process of the Nb[28.8 nm]/Au[3.9 nm]/Co[12.6 nm]/Au[4.4 nm] sample: (a) Nb[28.8 nm], (b) Au[3.9 nm], (c) Co[12.6 nm], and (d) Au[4.4 nm] surfaces. The direction of the incident electron beam is parallel to <1$\bar{1}$0> of the Nb(110) layer.

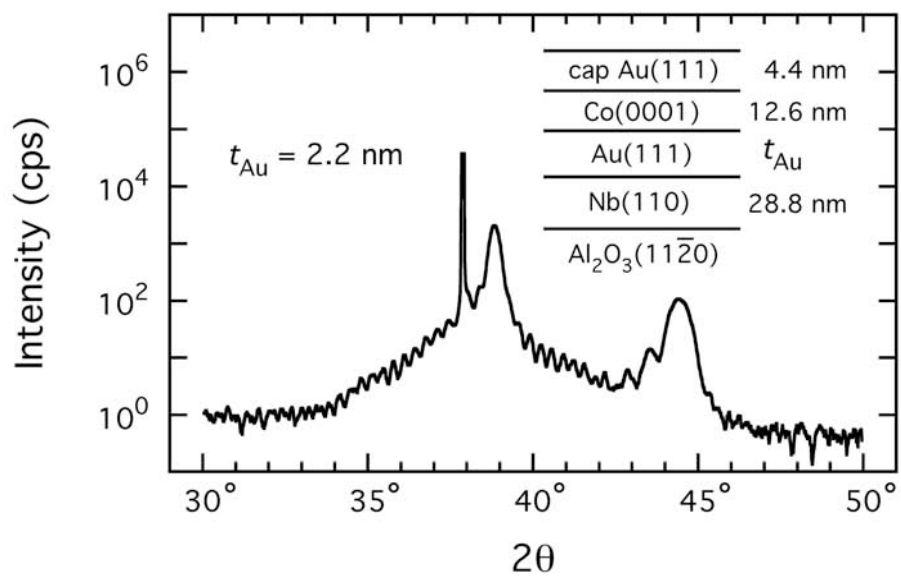

**FIG. 3.** Typical reflection x-ray diffraction pattern of middle-angle 2θ-θ scan for the $t_{Au}$=2.2 nm sample with Cu $K\alpha_1$ radiation. The intense peak at ~38° is from the $Al_2O_3(11\bar{2}0)$ substrate. Inset: schematic diagram of a vertical section of the sample structure and layer thicknesses.

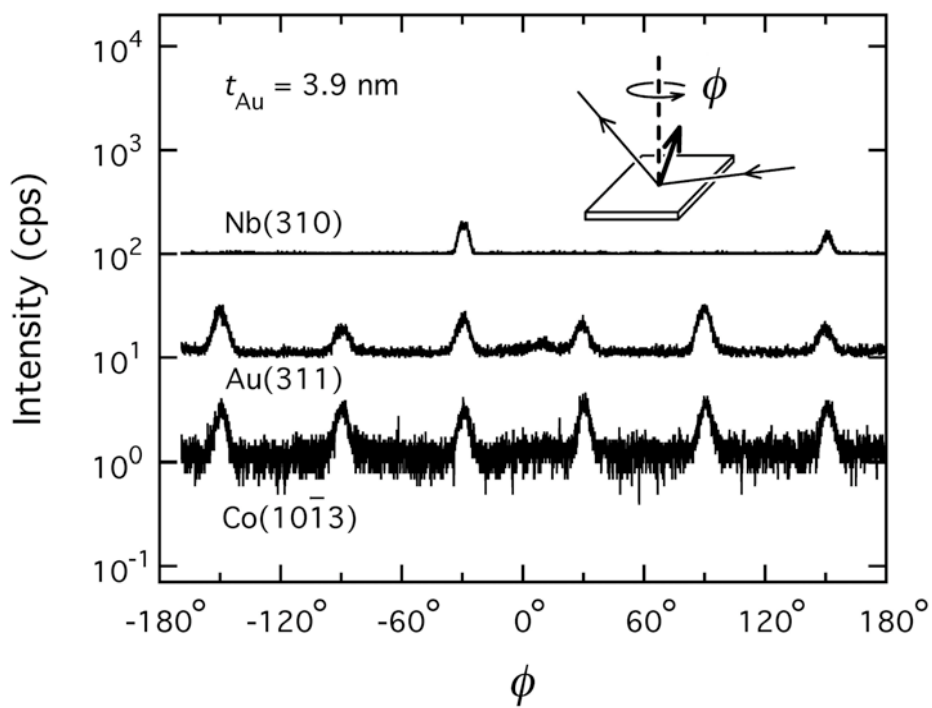

**FIG. 4.** X-ray diffraction patterns of $\phi$ scans for the $t_{Au}$=3.9 nm sample, where $\phi$ is the rotation angle around the axis perpendicular to the sample plane. Scans were carried out with the angles of $2\theta$ and $\omega$ fixed to the Nb(310), Au(311), and Co(10$\bar{1}$3) Bragg conditions. Inset shows the scattering geometry. Base lines are shifted arbitrarily for clarity of comparison. The direction of $\phi$=0° is parallel to <0001> of the $Al_2O_3$(11$\bar{2}$0) substrate.

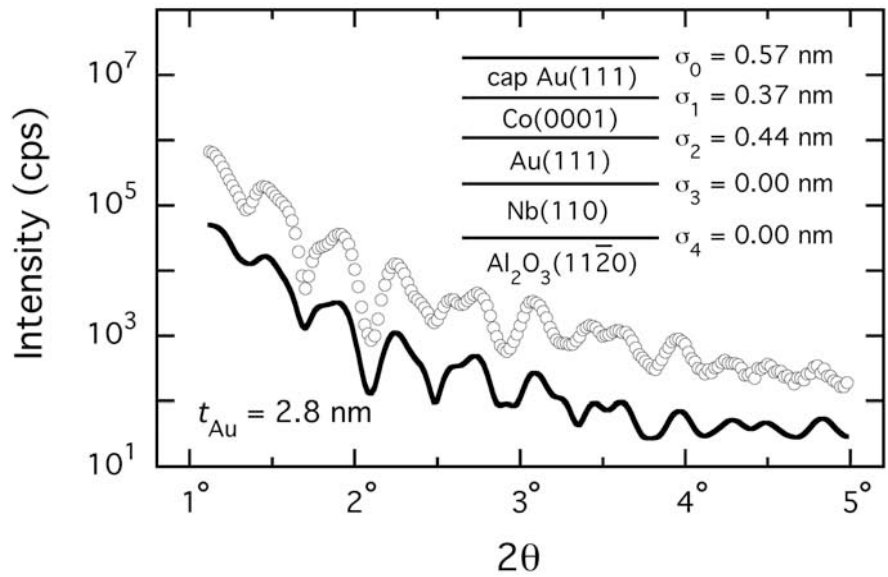

**FIG. 5.** Typical reflection x-ray diffraction pattern of small-angle $2\theta$-$\theta$ scan for the $t_{Au}$=2.8 nm sample (open circles). The solid curve corresponds to the optical-calculation result fitted to the experimental data using the profile-fitting program of SUPREX developed by Fullerton *et al.*[15] The calculation result is multiplied by 0.1 for clarity of comparison. Inset indicates the parameters of surface ($\sigma_0$) and interface [$\sigma_n$ ($n$=1-4)] roughness used in the calculation.

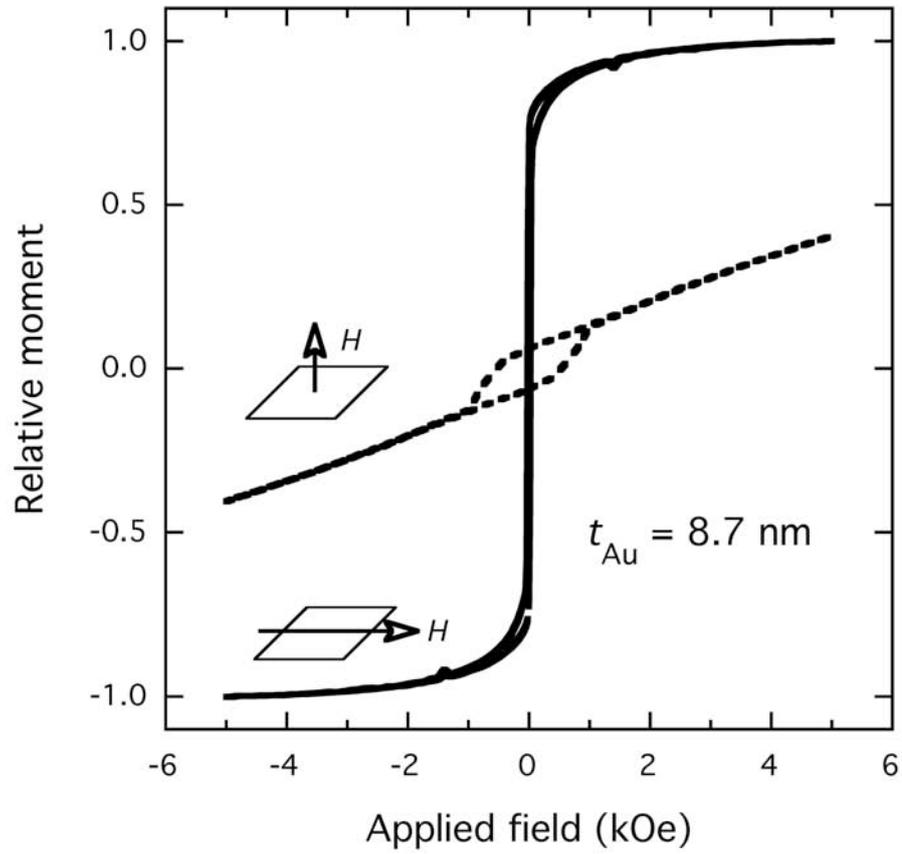

**FIG. 6.** Magnetic hysteresis loops measured at 8.0 K ($>T_c^{on}$=7.82 K) on a sample of $t_{Au}$=8.7 nm, with applied fields parallel (solid curve) and perpendicular (dashed curve) to the sample plane.

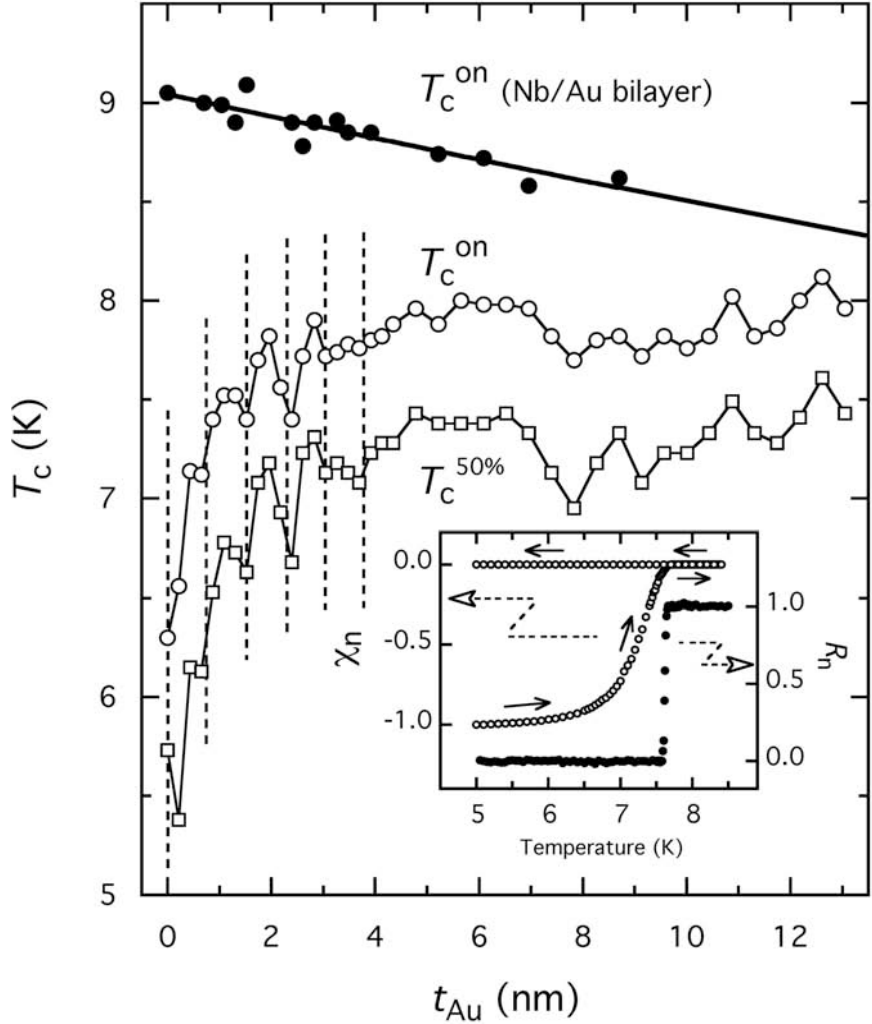

**FIG. 7.** Superconducting transition temperature $T_c$: $T_c^{on}$ (open circles) and $T_c^{50\%}$ (squares), as a function of $t_{Au}$ for the Nb[28.8 nm]/Au[$t_{Au}$]/Co[12.6 nm] trilayers. The data for the Nb[28.8 nm]/Au[$t_{Au}$] bilayers (filled circles) and its theoretical fit (solid curve) are also indicated for comparison.[7] The error of $T_c$ is within each symbol. The vertical broken lines are drawn at intervals of 0.76 nm (~3.2 ML of Au). Inset: typical temperature dependence of normalized magnetic susceptibility $\chi_n = \chi/|\chi_{(2K)}|$ at $H=2$ Oe and of normalized resistance $R_n = R/R_{(9K)}$ at $H=0$ Oe for the Nb/Au/Co trilayer of $t_{Au}=2.6$ nm, where $\chi_{(2K)}$ is the susceptibility at 2 K and $R_{(9K)}$ the resistance at 9 K.

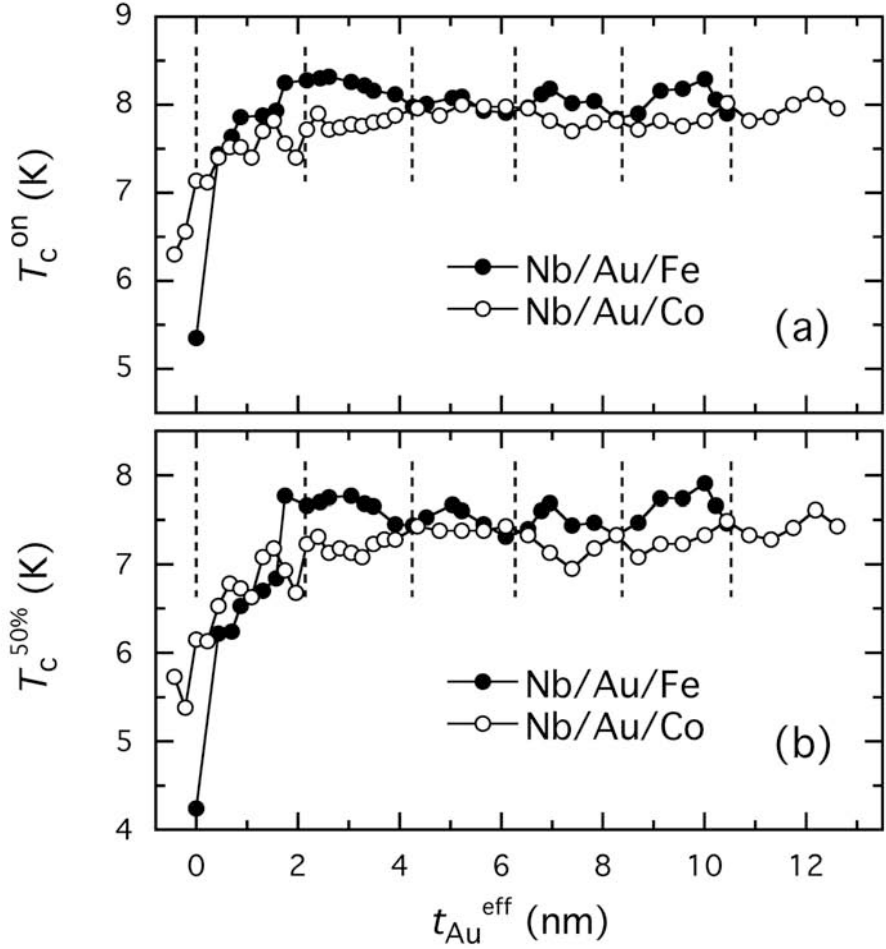

**FIG. 8.** Comparison between Nb[28.8 nm]/Au[$t_{Au}$]/Co[12.6 nm] and Nb[28.8 nm]/Au[$t_{Au}$]/Fe[12.6 nm] (Ref. 7) trilayers with respect to the dependence of (a) $T_c^{on}$ and (b) $T_c^{50\%}$ on $t_{Au}^{eff}$. Note that $t_{Au}^{eff}=t_{Au}$ for Nb/Au/Fe, while $t_{Au}^{eff}=t_{Au}-0.44$ nm for Nb/Au/Co in consideration of the interface-confined mixture at the Au/Co interface. The vertical broken lines are drawn at intervals of 2.1 nm (~9 ML).

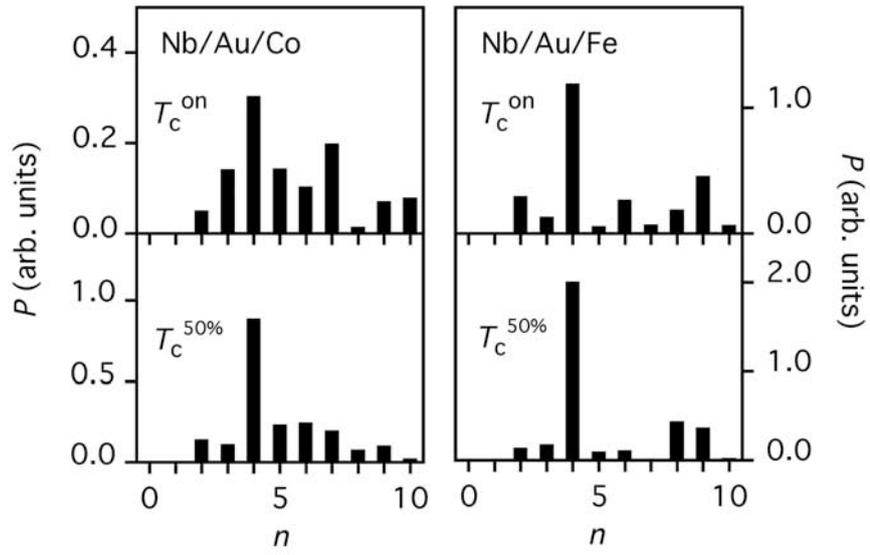

**FIG. 9** Power spectra of the $T_c^{on}(t_{Au}^{eff})$ and $T_c^{50\%}(t_{Au}^{eff})$ data as results of applying the FFT to the data of Nb/Au/Co (left panel) and Nb/Au/Fe (right panel). Only the positive spectra are shown. The horizontal scaling represents $n/(N \cdot \Delta t_{Au}^{eff})$, where $N(=20)$ is the number of data points and $\Delta t_{Au}^{eff}(=0.435$ nm) is an interval in $t_{Au}^{eff}$.

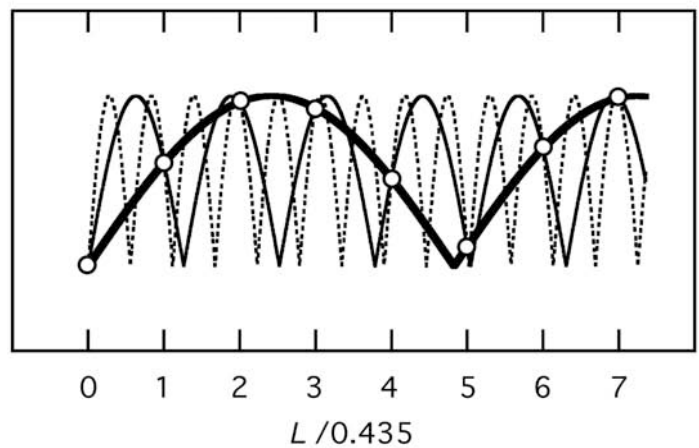

**FIG. 10** Mathematically possible aliasing effect. When sampled at discrete positions of $L=n{\times}0.435$ nm ($n$: integers), both the periodic functions with periods of 0.243 nm (thin broken curve) and 0.549 nm (thin solid curve) give a long-period oscillation with a period of 2.1 nm (thick solid curve)